\documentclass[amstex]{article}
\usepackage{graphicx}
\usepackage{dcolumn}
\usepackage{amsmath}
\usepackage{amssymb}
\usepackage{amsfonts}
\usepackage{amscd}
\begin{document}
\newtheorem{thm}{Theorem}[section]
\newtheorem{lem}{Lemma}[section]
\newtheorem{prop}{Proposition}[section]
\newtheorem{cor}{Corollary}[section]
\newtheorem{assum}{Assumption}[section]
\newtheorem{rem}{Remark}[section]
\newtheorem{defn}{Definition}[section]
\newcommand{\lv}{\left \vert}
\newcommand{\rv}{\right \vert}
\newcommand{\lV}{\left \Vert}
\newcommand{\rV}{\right \Vert}
\newcommand{\la}{\left \langle}
\newcommand{\ra}{\right \rangle}
\newcommand{\ket}[1]{\lv #1 \ra}
\newcommand{\bra}[1]{\la #1 \rv}
\newcommand{\lmk}{\left (}
\newcommand{\rmk}{\right )}
\newcommand{\al}{{\mathfrak A}}
\newcommand{\md}{M_d({\mathbb C})}
\newcommand{\ali}[1]{{\mathfrak A}_{[ #1 ,\infty)}}
\newcommand{\alm}[1]{{\mathfrak A}_{(-\infty, #1 ]}}
\newcommand{\cb}{{\cal B}}
\newcommand{\aax}{a(\alpha)}
\newcommand{\ana}{a_n(\alpha)}
\begin{center}
{\Large\bf Large Deviations in Quantum Spin Chain
} \\
\bigskip\bigskip
\bigskip\bigskip
\bigskip\bigskip
\bigskip\bigskip
{\Large Yoshiko Ogata}
\\
\bigskip
{\it 
Graduate School of Mathematics, Kyushu University
\\
1-10-6 Hakozaki, Fukuoka 812-8581, Japan}
\end{center}
\vfil
\noindent
\begin{abstract}
We show the
full large deviation principle for KMS-states
and $C^*$-finitely correlated states on a quantum spin chain.
We cover general local observables.
Our main tool is Ruelle's transfer operator method.
\end{abstract}

\noindent
\bigskip
\hrule
\bigskip
\section{Introduction}
While the large deviation for classical lattice
spin systems constitutes
a rather complete theory, our knowledge on
large deviations in quantum spin systems is still restricted.
Large deviation results for observables that depend only on
one site were established in high temperature KMS-states,
in \cite{nr}, using cluster expansion techniques.
In \cite{lr}, large deviation upper bounds were proven
for general observables, for KMS-states 
in the high temperature regime
and in dimension one.
Furthermore, it was shown that
a state in one dimension,
which satisfies
a certain factorization property
satisfies a large deviation upper bound \cite{hmo}.
This factorization property is satisfied
by KMS-states as well as $C^*$-finitely correlated states.
It was also shown in \cite{hmo}
that 
the distributions of the ergodic averages of
a one-site
observable with respect to an ergodic $C^*$-finitely 
correlated state satisfy full large deviation principle. 

In spite of these progresses, the theory in
quantum spin systems is not completed:
we do not know if the large deviation 
lower bound holds for general observables,
nor if the
large deviation upper bound holds in the intermediate temperature
KMS-states,
for more than two dimensional spin systems.
In this paper, we solve a part of the problem:
we prove the full large deviation principle
in dimension one.\\

The infinite spin chain with one site algebra 
${M_d}({\mathbb C})$ is given by the UHF $C^*$-algebra
\[
{\mathfrak A}_{{\mathbb Z}}:=\overline{
\bigotimes_{{\mathbb Z}}{M_d}({\mathbb C})}^{C^*},
\]
which is the $C^*$- inductive limit of the local algebras
\[
\left\{
{\mathfrak A}_{\Lambda}:=\bigotimes_{{\Lambda}}{M_d}({\mathbb C})\vert
\quad \Lambda\subset{\mathbb Z},\quad
| \Lambda|<\infty
\right\}
.
\]
For any subset $S$ of ${\mathbb Z}$,
we identify 
${\mathfrak A}_{S}:=
\overline{
\bigotimes_{{S}}{M_d}({\mathbb C})}^{C^*}$
with a subalgebra of ${\mathfrak A}_{{\mathbb Z}}$ under
the natural inclusion.
The algebra of local observables is defined by
\[
{\mathfrak A}_{loc}:= \cup_{| \Lambda |<\infty}{\mathfrak A}_{\Lambda}.
\]
Let $\gamma_j,\; j\in {\mathbb Z}$ be the $j$-lattice translation.
A state $\omega$ is called translation-invariant if 
$\omega\circ\gamma_j=\omega$ for all $j\in{\mathbb Z}$.
An interaction is a map $\Phi$ from 
the finite subsets
of $\mathbb Z$ into ${\mathfrak A}_{{\mathbb Z}}$ such
that $\Phi(X) \in {\mathfrak A}_{X}$ 
and $\Phi(X) = \Phi(X)^*$
for any finite $X \subset {\mathbb Z}$. 
In this paper, we will always assume that $\Phi$ is a finite range
translation-invariant interaction,
i.e., there exists $r\in {\mathbb N}$ such that
\[
\Phi(X)=0,\quad if \quad diam (X)>r,
\]
and $\Phi$ is invariant under $\gamma$,
\[
\Phi(X+j)=\gamma_j\lmk
\Phi(X)\rmk,\quad \forall j\in{\mathbb Z},\quad
\forall X\subset {\mathbb Z}.
\]
A norm of an interaction $\Phi$ is defined by
$\lV\Phi\rV\equiv\sum_{X\ni 0}\lv X\rv^{-1}\lV\Phi(X)\rV$.

For finite $\Lambda\subset{\mathbb Z}$, we set
\[
H_{\Phi}(\Lambda):=\sum_{I\subset{\Lambda}}\Phi(I).
\]
The distribution of 
$\frac{1}{n}H_{\Phi}([1,n])$ with respect to a state $\omega$
is the probability measure
\[
\mu_{n}(B):=\omega(1_{B}(\frac{1}{n}H_{\Phi}([1,n]))),\quad
B\in{\cal B},
\]
where ${\cal B}$ denotes the Borel sets of $\mathbb R$ and
$1_{B}(\frac{1}{n}H_{\Phi}([1,n]))\in\al_{[1,n]}$ is the spectral projection of
$\frac{1}{n}H_{\Phi}([1,n])$ corresponding to the set $B$.\\
Let $I:{\cal B}\to [0,\infty]$ be a lower semicontinuous mapping.
We say that we have
a large deviation upper bound for a closed set $C$
if
\begin{align*}
\lim_{n\to \infty}\sup\frac{1}{n}
\log\omega\lmk
1_{C}\lmk
\frac{1}{n}H_{\Phi}([1,n])\rmk
\rmk
\le -\inf_{x\in C} I(x).
\end{align*}
Similarly, 
we have 
a large deviation lower bound for an open set $O$
if
\begin{align*}
\lim_{n\to \infty}\inf\frac{1}{n}
\log\omega \lmk
1_{O}\lmk
\frac{1}{n}H_{\Phi}([1,n])\rmk
\rmk\ge -\inf_{x\in O}I(x).
\end{align*}
We say that
$\{\mu_n\}$ satisfies the (full) large deviation principle
if we have upper and lower bound for
all closed and open sets, respectively.
Furthermore, $I$ is said to be a good rate function if
all the level sets $\{x:I(x)\le \alpha\},\;\alpha\in[0,\infty)$
are compact subsets of ${\mathbb R}$( see \cite{dembo}).\\

In this paper, we show the full large deviation principle
for any kind of local observable, in
KMS-states and $C^*$-finitely correlated states
on quantum spin chain.
\paragraph{KMS-states}
Let $\Psi$ be a translation-invariant finite range interaction, 
and define
the finite volume 
Hamiltonian associated with a finite subset
$\Lambda \subset {\mathbb Z}$ by
\[
H_{\Psi}(\Lambda):=\sum_{I\subset{\Lambda}}\Psi(I).
\]
It is known that there exists 
a strongly continuous one parameter group of $*$-automorphisms
$\tau_{\Psi}$ on ${\mathfrak A}_{{\mathbb Z}}$, such that
\[
\lim_{\Lambda\nearrow{\mathbb Z}}
\lV
\tau_{\Psi}^t(A)-e^{itH_{\Psi}(\Lambda)}Ae^{-itH_{\Psi}(\Lambda)}
\rV
=0,\quad \forall t\in {\mathbb R},\quad\forall A\in{\mathfrak A}_{\mathbb Z}.
\]
The equilibrium state corresponding to the interaction $\Psi$
is characterized by the KMS condition. 
A state $\omega$
over ${\mathfrak A}_{\mathbb Z}$ is called
a $(\tau_{\Psi},\beta)$-KMS
state, if
\[
\omega(A\tau_{\Psi}^{i\beta}(B))=\omega(BA),
\]
holds for any pair $(A,B)$ of entire analytic elements
for $\tau_{\Psi}$.
It is known that one dimensional quantum spin system
has a unique $(\tau_{\Psi},\beta)$-KMS state for
all $\beta\in{\mathbb R}$ \cite{AraGib}.
In this paper, we prove the large deviation principle
for the $(\tau_{\Psi},\beta)$-KMS state:
\begin{thm}\label{tkms}
Let $\Psi$ be a translation-invariant finite range interaction and
 $\omega$ a $(\tau_{\Psi},\beta)$-KMS state.
Furthermore,
let $\Phi$ be another translation-invariant finite range interaction
and $\mu_{n,\Phi}$ the distribution of
$\frac{1}{n}H_{\Phi}([1,n])$ with respect to $\omega$.
Then the sequence $\{\mu_{n,\Phi}\}_{n\in {\mathbb N}}$ satisfies 
large deviation principle with a good rate function.
\end{thm}

\paragraph{Finitely correlated states}
The following recursive procedure to construct states on 
${\mathfrak A}_{\mathbb Z}$ was introduced in \cite{fc},
where the states obtained were called $C^*$-finitely correlated states.
For the construction one needs a triple $({\cal B},{\cal E},\rho)$,
where $\cal B$ is a finite dimensional $C^*$-algebra,
${\cal E} :M_d({\mathbb C})\otimes {\cal B}\to {\cal B}$
a unital completely positive map and $\rho$ a faithful state
on ${\cal B}$ with density operator $\hat\rho$.
Further, one has to assume that $\cal E$ and $\rho$
are related so that
$Tr_{\md}{\cal E}^*(\hat\rho)=\hat\rho$ holds.
Then
\[
\hat\varphi_1:={\cal E}^*(\hat\rho);\quad
\hat\varphi_n:=\lmk id_{\md}^{\otimes(n-1)}\otimes{\cal E}^*\rmk
\circ \cdots\circ\lmk id_{\md}\otimes{\cal E}^*\rmk\circ{\cal E}^*
\lmk\hat\rho\rmk;\; n=2,3,\cdots
\]
defines a state on 
${\md}^{\otimes n}\otimes{\cal B}$
for each $n\in {\mathbb N}$, 
and
\[
\hat\omega_n:=Tr_{\cal B}\hat\varphi_n
\]
gives a state $\omega_n$ on ${\md}^{\otimes n}$.
There exists a unique translation-invariant
state $\omega$ with local restrictions
$\omega\vert_{{\al}_{[1,n]}}=\omega_n$.
This is the $C^*$-finitely correlated state
generated by $(\cb,{\cal E},\rho)$.
In this paper, we prove large deviation principle
for $C^*$-finitely correlated states:
\begin{thm}\label{tfc}
Let $\omega$ be a $C^*$-finitely correlated state and
$\Phi$ a translation-invariant finite range interaction.
Let $\mu_{n,\Phi}$ be the distribution of
$\frac{1}{n}H_{\Phi}([1,n])$ with respect to $\omega$.
Then the sequence $\{\mu_{n,\Phi}\}_{n\in {\mathbb N}}$ satisfies 
large deviation principle with a good rate function.
\end{thm}
\paragraph{}
In order to study the large
deviations, we consider the corresponding
logarithmic moment generating function, defined by
\begin{align}\label{li}
f(\alpha)=\lim_{n\to\infty}\frac{1}{n}\log
\omega(e^{\alpha H_{\Phi}([1,n])}).
\end{align}

\begin{thm}[G{\"a}rtner-Ellis]
Let $\{\mu_n\}_{n\in{\mathbb N}}$ be a sequence of probability measures
on the Borel sets of $\mathbb R$.
Assume that the limit
\[
f(\alpha):=\lim_{n\to\infty}
\frac{1}{n}\log
\int e^{n\alpha x}d\mu_n(x)
\]
exists and is differentiable
for all $\alpha\in{\mathbb R}$.
Let
\[
I(x):=\sup_{\alpha\in {\mathbb R}}
\{
\alpha x-f(\alpha)
\}.
\]
Then $\{\mu_n\}$ satisfies the large deviation principle, i.e.,
we have
\begin{align*}
\lim_{n\to\infty}\sup \frac{1}{n}\log\mu_n(C)\le
-\inf_{x\in C}I(x),
\end{align*}
and 
\begin{align*}
\lim_{n\to\infty}\inf \frac{1}{n}\log\mu_n(O)\ge
-\inf_{x\in O}I(x),
\end{align*}
for any closed set $C$ and any open set $O$, respectively.
Furthermore, $I$ is a good rate function.
\end{thm}
In this paper, we use this Theorem to prove 
the large deviation principle, i.e., 
we prove the existence and differentiability of
the logarithmic moment generating function $f(\alpha)$
(\ref{li}).
Our main tool is the transfer
operator technique introduced 
by D.Ruelle for classical spin systems \cite{rue}.
H. Araki applied this method to quantum spin systems 
and showed the real analyticity of the mean
free energy \cite{AraGib}.
This paper is basically an extension of this result.
The non-commutative Ruelle transfer operator
was further generalized in \cite{gn} and \cite{matsui}.
We take advantage of these extensions.

The structure of this paper is as follows.
In Section \ref{ruelle}, we present a brief introduction to
the non-commutative Ruelle transfer operator technique.
In Section \ref{gibbs} and Section \ref{finite},
we prove
the large deviation principle,
for KMS-states and $C^*$-finitely correlated states,
respectively.
As a corollary of the result, 
we show the equivalence of ensembles, 
in Section \ref{eqe}.
\section{Non-commutative Ruelle transfer operator}\label{ruelle}
In this section, we give a brief introduction of
non-commutative Ruelle transfer operators
studied in 
\cite{AraGib} \cite {gn} and \cite{matsui}.
We represent a generalized form, 
but the Theorem \ref{mat} below can be proven in the same way as in
\cite{matsui}.
We follow the notation in \cite{matsui}
and consider one-sided infinite system $\al_{[1,\infty)}$.
We also introduce a finite dimensional $C^*$-algebra
$\cal B$.
By $Q^{(j)},\; j\in{\mathbb N}$, we denote the element of 
$1_{\cal B}\otimes \al_{[1,\infty)}$ with $Q$ in the $j$th component of
the tensor product of $\ali{1}$ and the unit in any other component.
Similarly, by $Q^{(0)}$ we denote an element in
${\cal B}\otimes{1_{\ali{1}}}$.
We introduce a $C^*$-algebra
\[
{\cal O}:=\lmk
{\cal B}\otimes \al_{[1,\infty)}\rmk
\otimes
\lmk
{\cal B}
\otimes\al_{[1,\infty)}
\rmk
\]
and consider automorphisms $\{\Theta_j\}_{j\in{\mathbb N}}$ 
of $\cal O$ determined by
\begin{align*}
\Theta_j\lmk Q^{(k)}\otimes 1\rmk=\left\{
\begin{gathered}
1\otimes Q^{(k)},\quad for\; k\ge j\\
Q^{(k)}\otimes 1, \quad for\; k<j,
\end{gathered}
\right.\\
\Theta_j\lmk 1 \otimes Q^{(k)}\rmk=\left\{
\begin{gathered}
Q^{(k)}\otimes 1,\quad for\; k\ge j\\
1\otimes Q^{(k)}, \quad for\; k<j
\end{gathered}
\right.
.
\end{align*}
For any element $Q$ in 
${\cal B}\otimes{\ali{1}}
$, 
we set 
\begin{align*}
{\rm var}_j(Q):=\lV
\Theta_j(Q\otimes 1)-Q\otimes 1
\rV,\quad j\in{\mathbb N}.
\end{align*}
For any $\theta$ satisfying $0<\theta<1$ and 
$Q\in \cb\otimes{\ali{1}}$,
we set
\begin{align*}
\lV Q\rV_\theta:=\max
\left\{
\frac{{\rm var}_j Q}{\theta^j},\quad j\in{\mathbb N}
\right\}.
\end{align*}
By $F_\theta$ we denote the dense subalgebra
of $\cb\otimes{\ali{1}}$ consisting of elements $Q$ with finite $\lV Q\rV_\theta$,
and introduce the norm $\lV| Q|\rV$ of $F_\theta$
via the following equation:
\begin{align*}
\lV| Q|\rV=\max\{\lV Q\rV,\lV Q\rV_\theta\}.
\end{align*}
$F_\theta$ is complete in this norm.\\
We need the $*$-isomorphism $\tau_{c+}$, (resp. $\tau_{c-}$)
of ${\cal B}\otimes \al_{[2,\infty)}\backsimeq 
{\cal B}\otimes1_{\al_{\{1\}}}\otimes \al_{[2,\infty)}$
onto ${\cal B}\otimes \al_{[1,\infty)}$
(resp. $\al_{(-\infty, -2]}\otimes{1_{ \al_{\{-1\}}}}\otimes
{\cal B}
\backsimeq \al_{(-\infty, -2]}\otimes {\cal B}$
onto $\al_{(-\infty, -1]}\otimes {\cal B}$)
determined by
\begin{align*}
\tau_{c+}\lmk
x\otimes id_{\al{\{1\}}}\otimes y
\rmk
=x\otimes y,
\end{align*}for all $x\in {\cal B}$
and $y\in\al_{[1,\infty)}$,
(resp.
\begin{align*}
\tau_{c-}\lmk
y\otimes id_{\al{\{-1\}}}\otimes x
\rmk
=y\otimes x,
\end{align*}
for all $x\in {\cal B}$ and $y\in\al_{(-\infty, -1]}$.)\\

We now introduce a Ruelle transfer operator $L$:
\begin{assum}\label{one}
Let $a$ be an element in ${\cal B}\otimes \ali{1}$,
and
\[
{\cal E} :\cb\otimes {\md}\to{\cal B}
\]
a completely positive unital map.
Define 
a Ruelle transfer operator $L$ on $\cb\otimes \ali{1}$
by
\begin{align}\label{rop}
L(Q):=
\tau_{c,+}\lmk {\cal E}\otimes id_{[2,\infty)}\rmk
(a^*Qa) 
,\quad
Q\in \cb\otimes \ali{1}.
\end{align}
Assume that 
\begin{description}
\item[(i)]The element $a$ is in $F_\theta$ and 
invertible in $F_\theta$.
\item[(ii)]There exists an invariant state $\varphi$
of $L$.
\item[(iii)]There exists a positive constant $K$ such that
the following bound is valid:
Let $Q$ be any strictly positive element in 
$\cb\otimes(\al_{loc}\cap\ali{1})$.
There exists a positive integer $N=N(Q)$ satisfying
\[
L^n(Q)\le K\inf L^n(Q),\quad \forall n\ge N.
\]
\end{description}
\end{assum}
If Assumption \ref{one} is valid, the restriction
of $L$ to the Banach space $F_\theta$ gives a bounded operator
on $F_\theta$.
Assumption \ref{one} guarantees the following properties of $L$.
\begin{thm}\label{mat}
Let $L$ be a Ruelle transfer operator satisfying Assumption \ref{one}.
Then
\begin{description}
\item[(i)]There exists an element $h$ in $F_\theta$
and a positive constant $m>0$ such that
\[
L(h)=h,\quad m\le h,\quad \varphi(h)=1.
\]
\item[(ii)]
Define an operator $L_h$ and a state $\varphi_h$ by
\[
L_h(Q):=h^{-\frac 12}L\lmk h^{\frac 12}Q h^{\frac 12}\rmk
h^{-\frac 12},\quad Q\in\cb\otimes\ali{1},
\]
and
\[
\varphi_h(Q):=\frac{\varphi\lmk h^{\frac 12}Q 
h^{\frac 12}\rmk}{\varphi(h)},\quad Q\in\cb\otimes\ali{1}.
\]
Then $L_h$ gives a bounded operator on the Banach space 
$F_\theta$ and
there exists $\delta_1>0$ and $C_1>0$
such that
\begin{align}\label{eb}
\vert\Vert L^n_h(Q)-\varphi_h(Q)\Vert\vert
\le C_1 e^{-\delta_1 n}\vert\Vert Q\Vert\vert,
\end{align}
for all $n\in{\mathbb N}$ and $Q\in F_\theta$.
\item[(iii)]$\lim_{n\to\infty}
\Vert L^n(1)-h\Vert=0.$
\item[(iv)]
As a bounded operator on $F_\theta$, $L$ has a nondegenerate eigenvalue
$1$ and rest of the spectrum has modulus less than $e^{-\frac {\delta_1} {2}}$.
\end{description}
\end{thm}
{\it Proof}
The proof is completely analogous to that in 
\cite{matsui}.
We omit the details.
$\square$\\

Now we consider a family of Ruelle transfer operators
$\{L_\alpha\}_{\alpha\in {\mathbb R}}$.
\begin{thm}\label{spec}
Let $\{L_\alpha\}_{\alpha\in {\mathbb R}}$
be a family of operators on $\cb\otimes \ali{1}$.
Suppose that
each $L_\alpha$ is of the form 
(\ref{rop}) with $a=a(\alpha)\in \ali{1}$ and
${\cal E}: \cb \otimes \md\to \cb$,
satisfying
 (i), (iii) of Assumption \ref{one}.
Assume that
the map
\[
{\mathbb R}\ni \alpha \mapsto L_\alpha\in B(F_\theta)
\]
has a $B(F_\theta)$-valued analytic extension to a neighborhood
of $\mathbb R$.
Then, as a bounded operator on $F_\theta$, 
each $L_\alpha$ 
has a strictly positive nondegenerate eigenvalue
$\lambda(\alpha)$
such that
\begin{description}
\item[(i)]
$\lambda(\alpha)$ has a
strictly positive eigenvector $h(\alpha)\in F_\theta$,
and
\[
\lim_{n\to\infty}\lV
\lambda(\alpha)^{-n}L_\alpha^n(1)-h(\alpha)
\rV=0,
\]
\item[(ii)]
${\mathbb R}\ni \alpha\mapsto \lambda(\alpha)$ is differentiable.
\end{description}
\end{thm}
\begin{rem}
An analogous result for left-side chain $\alm{-1}$ holds.
\end{rem}
{\it Proof}\\
There exists a state $\varphi_\alpha$
and a strictly positive scalar $\lambda(\alpha)$ such that
$L_\alpha^*\varphi_\alpha=\lambda(\alpha)\varphi_\alpha$.
In fact, by the invertibility of $a(\alpha)$ and unitality of $\cal E$,
we have
\[
L_\alpha(1)\ge \lV a(\alpha)^{-1}\rV^{-2}>0.
\]
Accordingly, if $\nu$ is a state of $\cb\otimes \ali{1}$,
a state
\[
G(\nu)(Q):=\frac{\nu(L_\alpha(Q))}{\nu(L_\alpha(1))},\quad
Q\in \cb\otimes \ali{1}
\]
is well defined.
This map $G$ is weak$^*$continuous on the state space.
Therefore, using Schaudar Tychonov theorem,
we can show the existence of a fixed point of $G$,
i.e.,  a state $\varphi_\alpha$
and a strictly positive scalar $\lambda(\alpha)$ such that
$L_\alpha^*\varphi_\alpha=\lambda(\alpha)\varphi_\alpha$.
(See \cite{AraGib}).

The operator $\lambda(\alpha)^{-1}L_{\alpha}$ 
satisfies Assumption \ref{one}.
Applying Theorem \ref{mat} to $\lambda(\alpha)^{-1}L_{\alpha}$,
we obtain (i).
By (iv) of Theorem \ref{mat} and 
regular perturbation theory,
differentiability of $\lambda(\alpha)$ can be proven.
$\square$\\

We will construct Ruelle operators $L_\alpha$ 
so that the eigenvalue $\lambda(\alpha)$ in Theorem \ref{spec}
corresponds to the logarithmic moment generating function $f(\alpha)$
in (\ref{li}).
\section{Large deviation principle for KMS-states}\label{gibbs}
Let $\Psi$ be a finite range interaction and
$\omega$ a
unique $(\tau_\Psi,\beta)$-KMS state.
Let $\Phi$ be another finite range interaction.
In this section, we prove large deviation principle 
of the distribution of $\frac 1n H_\Phi([1,n])$
in $\omega$, Theorem \ref{tkms}.
By the G{\"a}rtner-Ellis Theorem, it suffices to show 
the existence and 
differentiability of the logarithmic moment generating function
\begin{align}\label{mgf}
f(\alpha)=\lim_{n\to\infty}\frac{1}{n}\log
\omega(e^{\alpha H_{\Phi}([1,n])}),\quad \forall\alpha\in {\mathbb R}.
\end{align}
\begin{lem}\label{pn}
Let $p_n(\alpha)$ be
\[
p_n(\alpha):=Tr_{[1,n]}
\lmk
e^{-\frac{\beta}{2}H_\Psi[1,n]} e^{\alpha H_\Phi[1,n]}
e^{-\frac{\beta}{2}H_\Psi[1,n]}
\rmk,\quad \alpha\in {\mathbb R}.
\]
It suffices to prove
the existence and differentiability of the 
limit
\begin{align}\label{pl}
\lim_{n\to\infty}\frac{1}{n}\log
p_n(\alpha),\quad \forall \alpha\in{\mathbb R}.
\end{align}
\end{lem}
{\it Proof} \;
In \cite{lr}, it was shown that
there exists a positive constant
$C_1$ such that
\begin{align}\label{st}
C_1^{-1}\omega_n\le \omega\vert_{\al_{[1,n]}}
\le C_1 \omega_n,
\end{align}
where
$\omega_n$ is a state on $\al_{[1,n]}$
given by
\[
\omega_n(A)=\frac{Tr_{[1,n]} e^{-\beta H_{\Psi}{[1,n]} }A}
{Tr_{[1,n]} e^{-\beta H_{\Psi}{[1,n]}}}.
\]
From this inequality, we have
\begin{align*}
&\lim_{n\to\infty}\frac{1}{n}
\lmk
\log p_n(\alpha)-\log\omega(e^{\alpha H_\Phi{[1,n]}})
-\log Tr_{[1,n]}e^{-\beta H_{\Psi}{[1,n]}}
\rmk\\
&=
\lim_{n\to\infty}\frac{1}{n}
\lmk
\log \omega_n(e^{\alpha H_\Phi{[1,n]}})-\log\omega(e^{\alpha H_\Phi{[1,n]}})
\rmk
=0.
\end{align*}
As the existence of the limit
\[
\lim_{n\to\infty}\frac{1}{n}\log Tr_{[1,n]} e^{-\beta H_{\Psi}{[1,n]}}
\]
is known, it suffices to prove
the existence and differentiability of the 
limit (\ref{pl}).$\square$\\

By Lemma \ref{pn}, we shall confine our attention to the analysis of
$p_n(\alpha)$.
We will freely use the notations in Appendix \ref{anal}.\\

We now define a family of Ruelle transfer operators 
$\{L_\alpha\}_{\alpha\in{\mathbb R}}$,
given in the form of (\ref{rop}):
we set $\cb=\md$,
and define a completely positive unital map
${\cal E}: \md\otimes\md\to \md$,
through the formula ${\cal E}(a\otimes b):=d^{-1}Tr_{\md}(a)b$.
Furthermore, for each $\alpha\in{\mathbb R}$,
 we define $a(\alpha)$ by
\begin{align*}
a(\alpha):=\tau_{\Phi_{[1,\infty)}}^{-i\frac{\alpha}{2}}
\lmk
E_r
\lmk
-\frac{\beta}{2}\hat H_\Psi^r(1);
-\frac{\beta}{2}H_\Psi[2,\infty)
\rmk
\rmk
E_r
\lmk
\frac{\alpha}{2}\hat H_\Phi^r(1);
\frac{\alpha}{2}H_\Phi[2,\infty)
\rmk\in
\ali{1}.
\end{align*}
The Ruelle transfer operator on $\cb\otimes\ali{1}=\ali{0}$ 
is given by
\begin{align}
L_\alpha(Q):=\gamma_{-1}
\lmk
d^{-1}Tr_{\{0\}}\otimes id_{[1,\infty)}
\rmk
\lmk
{a(\alpha)}^*Qa(\alpha)
\rmk.
\end{align}

In order to apply Theorem \ref{spec}, we have to check
that each $L_\alpha$ satisfies the Assumption \ref{one},
(i) and (iii) :
\begin{lem}\label{asok}
Each $L_\alpha,\; \alpha\in{\mathbb R}$ 
satisfies the Assumption \ref{one}, (i),(iii).
\end{lem}
{\it Proof}\;
It was shown in \cite{AraGib}
that for any local element $Q$ in $\ali{1}$,
a subset $I\subset [1,\infty)$,
and a finite range interaction $\Phi$,
$E_r(Q; H_\Phi(I))$
is an invertible element in $\al_1$, which is the subalgebra
of $F_\theta$ defined by (\ref{a1}).
Furthermore, 
any element in $\al_1$ 
is entire analytic for $\tau_{\Phi_I}$,
and $\tau_{\Phi_I}$ acts on $\al_1$ as a group of automorphisms
with one complex parameter.(See Appendix \ref{anal}.)
Therefore, 
$a(\alpha)$ belongs to $\al_1$
and invertible in $\al_1$, so
$a(\alpha)$ belongs to $F_\theta$
and invertible in $F_\theta$.
Hence, (i) of Assumption \ref{one} is satisfied.

The proof of (iii) goes parallel to the argument in \cite{matsui},
where an example of Ruelle transfer operator was considered.
We shall first
write $L^n_\alpha$
in a more tractable form.
By an inductive calculation,
we obtain
\begin{align*}
L_\alpha^n(Q)= d^{-n}\gamma_{-n}\circ
\lmk
Tr_{[0,n-1]}
\otimes id_{[n,\infty)}
\rmk\circ 
\lmk
\tilde a_n^*(\alpha) Q\tilde a_n(\alpha)
\rmk,
\end{align*}
where 
we denoted 
$\aax\gamma_1(\aax)\gamma_2(\aax)\cdots
\gamma_{(n-1)}(\aax)$ by $\tilde a_n(\alpha)$.
It is not hard to prove
\begin{align*}
\tilde a_n(\alpha)=
\tau_{\Phi_{[1,\infty)}}^{-i\frac{\alpha}{2}}
\lmk E_r\lmk -\frac{\beta}{2} \hat H^r_\Psi(n);
-\frac{\beta}{2} H_\Psi([n+1,\infty)\rmk\rmk
 E_r\lmk \frac{\alpha}{2} 
\hat H_\Phi^r(n);\frac{\alpha}{2} H_\Phi([n+1,\infty)\rmk,
\end{align*}
using (\ref{com}).
Let $a_n(\alpha),\; n\ge 2$ be
\begin{align*}
a_n(\alpha):=\tilde a_n(\alpha)e^{\frac{\beta}{2} H_{\Psi}[1,n-1]}
e^{-\frac{\alpha}{2} H_{\Phi}[1,n-1]}.
\end{align*}
Using the relation (\ref{com}) again, we can show 
\begin{align*}
a_n(\alpha)=&
\tau_{\Phi_{[1,\infty)}}^{-i\frac{\alpha}{2}}
\lmk E_r\lmk -\frac{\beta}{2} 
 W^r_\Psi(n);-\frac{\beta}{2} \lmk
H_\Psi([1,n-1])
+H_{\Psi}[n+1,\infty\rmk)\rmk\rmk\\
& E_r\lmk \frac{\alpha}{2} W^r_\Phi(n);\frac{\alpha}{2} 
\lmk
H_\Phi[1,n-1]+H_\Phi([n+1,\infty)
\rmk
\rmk.
\end{align*}
Furthermore, we define a completely positive unital map
$\varphi_n :\ali{0}\to\ali{0},\; n\ge 2$, 
by
\begin{align*}
\varphi_n(Q):=& p_{n-1}^{-1}(\alpha)d^{-1}
\gamma_{-n}\circ
\lmk
Tr_{[0,n-1]}\otimes id_{[n,\infty)}
\rmk\\
&\lmk
e^{-\frac{\beta}{2}H_\Psi[1,n-1]} e^{\frac{\alpha}{2}H_\Phi[1,n-1]}
Qe^{\frac{\alpha}{2}H_\Phi[1,n-1]}e^{-\frac{\beta}{2}H_\Psi[1,n-1]}
\rmk.
\end{align*}
Using these notations, we can rewrite $L_\alpha^n$ as
\begin{align}\label{Lrw}
L_\alpha^n(Q)= d^{-(n-1)}p_{n-1}(\alpha) \varphi_n(\ana^* Q\ana),\quad
n\ge 2.
\end{align}

Next we evaluate (\ref{Lrw}), using the properties of $\ana$
given in 
Lemma \ref{ks}:
that is, \begin{align}\label{qc}
\lim_{n\to\infty}\lV[ Q, \ana]\rV=0,\quad \forall
Q\in \al_{loc},
\end{align}
and that
there exists a positive constant $C$ such that
\begin{align}\label{ku}
&
\sup_{n\in {\mathbb N}}
\lV a_n(\alpha)\rV,\;
\sup_{n\in {\mathbb N}}
\lV \lmk a_n(\alpha)\rmk^{-1}\rV<C.
\end{align}
Let $Q$ be any strictly positive element in $\al_{[0,n_0]}$.
By (\ref{qc}), we
can choose $\varepsilon>0$ and $N(Q)\in{\mathbb N}$ so that
\[
4C^3\lV Q^{\frac 12}\rV \varepsilon \le\inf Q,
\]
and
\begin{align*}
N(Q)\ge n_0+1,\quad
\lV
[Q^{\frac 12},\ana]
\rV<\varepsilon,\;
\forall n\ge N(Q).
\end{align*}
As $\varphi_n$ is a completely positive unital map,
we have $\lV\varphi_n\rV=\lV\varphi_n(1)\rV=1$.
Note that $\varphi_n(Q)$
is a scalar if $n-1\ge n_0$.
Thus we get
\begin{align*}
&L_\alpha^n(Q)\le
d^{-(n-1)} p_{n-1}(\alpha)\lmk
C^2 \varphi_n(Q)+2C\lV Q^{\frac12}\rV\lV[Q^{\frac12},\ana]\rV
\rmk\\
&\le d^{-(n-1)}
p_{n-1}(\alpha)\lmk\frac{1}{2C^2}+C^2\rmk\varphi_n(Q),
\end{align*}
and
\begin{align*}
&L_\alpha^n(Q)\ge d^{-(n-1)}
p_{n-1}(\alpha)\lmk
-2C\lV Q^{\frac 12}\rV
\lV[Q^{\frac12},\ana]\rV
+\frac{1}{C^2}\varphi_n(Q)
\rmk\\
&\ge d^{-(n-1)}p_{n-1}(\alpha)
\frac{1}{2C^2}\varphi_n(Q),
\end{align*}
for all $n\ge N(Q).$
Hence we obtain (iii) of Assumption \ref{one}:
\[
L_\alpha^n(Q)\le
(1+2C^4)\inf L_\alpha^n (Q),
\]
for all $n\ge N(Q)$.
$\square$\\\\
{\it Proof of Theorem \ref{tkms}}\\
Note that
\[
{\mathbb R}\ni\alpha\mapsto
L_\alpha\in B(F_\theta)
\]
has a $B(F_\theta)$-valued analytic extension to 
a neighborhood
of $\mathbb R$.
We thus can apply Theorem \ref{spec}
to $\{L_\alpha\}$.
Accordingly, each $L_\alpha$ has a strictly positive 
eigenvalue $\lambda(\alpha)$ associated with 
a strictly positive eigenvector $h(\alpha)$
such that 
\[
\lim_{n\to\infty}\lV
\lambda(\alpha)^{-n}L_\alpha^n(1)-h(\alpha)
\rV=0.
\]
Furthermore, ${\mathbb R}\ni \alpha\mapsto \lambda(\alpha)$
is differentiable.
By (\ref{Lrw}) and (\ref{ku}),
we have
\begin{align}
d^{-(n-1)}p_{n-1}(\alpha)C^{-2}
\le L_\alpha^n(1)= d^{-(n-1)}p_{n-1}(\alpha) \varphi_n(\ana^* 1\ana)
\le d^{-(n-1)}
p_{n-1}(\alpha) C^{2}.
\end{align}
Hence for any state $\nu$ on $\ali{0}$, we have
\begin{align*}
&\lim_{n\to\infty}\frac{1}{n-1}
\lmk\log p_{n-1}(\alpha)-\log \nu(\lambda(\alpha)^{-n}L_\alpha^n(1))-n\log\lambda(\alpha)
-(n-1)\log d
\rmk\\
&=\lim_{n\to\infty}\frac{1}{n}
\lmk\log p_n(\alpha)\rmk
-\log\lambda(\alpha)-\log d
=0.
\end{align*}
Therefore, the limit
\begin{align}
\lim_{n\to\infty}\frac{1}{n}\log
p_n(\alpha)=\log\lambda(\alpha)+\log d
,\quad \forall \alpha\in{\mathbb R}.
\end{align}
exists and is differentiable.
Applying Lemma \ref{pn}, we have thus proved the Theorem.
$\square$

\section{Large deviation principle for $C^*$-finitely correlated states}\label{finite}In this section, we prove the large deviation principle for 
finitely correlated states, Theorem \ref{tfc}.
Let $\omega$ be a $C^*$-finitely correlated state
generated by a finite dimensional
$C^*$-algebra $\cb$,
a completely positive unital map 
${\cal E} :\md\otimes\cb\to\cb$ and a faithful state $\rho$.
By the translation invariance of $\omega$, it suffices to 
show that the limit
\begin{align}\label{fl}
\lim_{n\to\infty}\frac 1n \log\omega\left(
e^{\alpha H_\Phi[-n,-1]}
\right)
\end{align}
exists and is differentiable.
We define a completely positive unital map
$\hat {\cal E}_{1}:{\cal B}\to{\cal B}$
through the formula 
$\hat {\cal E}_{1}(b):={\cal E}({1}\otimes b),\; b\in{\cal B}$.
\begin{lem}\label{red}
It suffices to show the existence and differentiability 
of the limit
(\ref{fl}) for
$\omega$ generated by a triple $(\cb,{\cal E},\rho)$
satisfying the following condition:
there exists 
a positive constant $s>0$ such that
\begin{align}\label{cpb}
s^{-1}\rho(b)\le\lmk
\hat{\cal E}_{1}\rmk(b)\le s\rho(b),\quad
0\le \forall b,\; b\in {\cal B}.
\end{align}
\end{lem}
{\it Proof} \;
It is known that every $C^*$-finitely correlated state 
has a unique decomposition as a {\it finite} convex combination of extremal
periodic states, which are again $C^*$-finitely correlated \cite{fc}.
That is, we can write $\omega$ as a finite sum
$\omega=\sum_{i=1}^n\lambda_i\omega_i,\; 0< \lambda_i,\; \sum_{i=1}^n\lambda_i=1$,
where each $\omega_i$ is an extremal $p_i$ periodic
state.
Furthermore, $\omega_i$ is a $C^*$-finitely correlated state
on $(M_d({\mathbb C})^{\otimes p_i})_{\mathbb Z}$,
generated by a triple $(\cb_i,{\cal E}_i,\rho_i)$,
such that 
$1$ is the only eigenvector of $(\hat{\cal E}_i)_1$
with eigenvalue one,
and rest of the spectrum 
has modulus strictly less than $1$.
Therefore, 
it suffices 
to consider
$\omega$
generated by a completely positive map
${\cal E}$ such that $\hat{\cal E}_1$
has a nondegenerate eigenvalue
$1$ and rest of the spectrum 
has modulus strictly less than $1$.
We shall confine our attention to this case.

Next we claim that 
there exists an integer $l$ and a positive constant $s>0$ such that
\begin{align}
s^{-1}\rho(b)\le\lmk
\hat{\cal E}_{1}\rmk^l(b)\le s\rho(b),\quad
0\le b,\; b\in {\cal B}.
\end{align}
To see this, let $P$ be a spectral
projection of $\hat{\cal E}_1$ 
corresponding to
the eigenvalue $1$,
and set $\bar P=1-P$.
By assumption, the range of $P$ is ${\mathbb C}1$.
As $\rho$ is a faithful state 
on a finite dimensional $C^*$-algebra,
there exists $c>0$ such that $\hat \rho\ge c 1$.
Accordingly, we have $c\lV b\rV\le \rho(b),\;\forall b\ge 0,b\in \cb$.
By the assumption, if we take $l$ large enough,
we have
\[
\lV
(\hat{\cal E}_{1})^l\bar P(b)
\rV\le\frac{c}{2}
\lV b\rV,\quad\forall b\in{\cal B}.
\]
Furthermore, we have
\[
\rho(b)=\lim_{n\to\infty}\rho\lmk
\hat{\cal E}_1^n(b)\rmk
=\rho(P(b)).
\]
We thus obtain the claim:
there exists $l$ such that
\begin{align*}
\frac{1}{2}\rho(b)\le
\rho(b)-\frac{c}{2}
\lV b\rV
\le\hat
{\cal E}^l_{1}(b)=\hat{\cal E}^l_{1}(Pb)+\hat{\cal E}^l_{1}(\bar Pb)
=\rho(b)+\hat{\cal E}^l_1\lmk
\bar P(b)
\rmk
\le
\rho(b)+\frac{c}{2}
\lV b\rV
\le\frac{3}{2}\rho(b),
\end{align*}
for $0\le b,\; b\in\cb$.

Note that $\omega$ is a $C^*$-finitely correlated state
on $((\md)^{\otimes l})_{\mathbb Z}$,
generated by
$(\cb, {\cal E}^{(l)}, \rho)$,
where ${\cal E}^{(l)}$ is the $l$-th iterate of $\cal E$.
Furthermore, we have
$\hat{{\cal E}^{(l)}}_1=(\hat {\cal E}_1)^l$.
Therefore,
it suffices to consider
$\omega$ generated by a triple $(\cb,{\cal E},\rho)$
satisfying (\ref{cpb}).
$\square$\\

We shall confine our attention to $\omega$ satisfying (\ref{cpb}).\\

As a transfer operator, we consider a map from
$\alm{-1}\otimes{\cal B}$ to $\alm{-1}\otimes{\cal B}$.
For each $\alpha\in {\mathbb R}$, we define 
$L_\alpha$ by
\begin{align*}
L_\alpha(Q):=
\tau_{c-}\circ\lmk
id_{(-\infty,-2]}\otimes{\cal E}
\rmk
\lmk
a(\alpha)^* Q a(\alpha)
\rmk,\quad Q\in \alm{-1}\otimes\cb
\end{align*}
Here,
$a(\alpha)$
is an element of $\alm{-1}$ given by
\begin{align*}
a(\alpha):=E_r\lmk
\frac{\alpha}{2} \hat H_{\Phi}^l(-1);
\frac{\alpha}{2} H_\Phi(-\infty, -2]
\rmk.
\end{align*}
\begin{lem}\label{asokf}
Each $L_\alpha,\; \alpha\in{\mathbb R}$ 
satisfies (i), (iii) of Assumption \ref{one}.
\end{lem}
{\it Proof}\;
As in Section \ref{gibbs}, $a(\alpha)$ is an invertible element of $F_\theta$
and (i) holds.\\
We prove (iii).We shall first
write $L^n_\alpha$
in a more tractable form.
By an inductive calculation, we obtain
\[
L_\alpha^n(Q)=
\lmk
\tau_{c-}\circ(id_{(-\infty,-2]}\otimes{\cal E})
\rmk^n
\lmk
\tilde a_n(\alpha)^* Q\tilde a_n(\alpha)
\rmk,
\]
where
\[
\tilde a_n(\alpha):=\aax\gamma_{-1}(\aax)\cdots\gamma_{-(n-1)}(\aax).
\]
Let $\ana, n\ge 2$ be
\[
\ana:=\tilde a_n(\alpha) e^{-\frac{\alpha}{2}H_\Phi[-n+1,-1]}.
\]
For each $n\ge 2$, we define a positive constant $p_n(\alpha)$, 
a completely positive map $\Phi_n$
by
\begin{align}\label{ndef}
&p_n(\alpha):=\omega\lmk
e^{\alpha H_\Phi[-n+1,-1]}\rmk\nonumber
\\
&\Phi_n(Q):=
p_n^{-1}(\alpha)
\lmk
\tau_{c-}\circ(id_{(-\infty,-2]}\otimes{\cal E})
\rmk^n
\lmk
e^{\frac\alpha 2 H_\Phi[-n+1,-1]}Q
e^{\frac\alpha 2 H_\Phi[-n+1,-1]}
\rmk.\nonumber\\
&
\end{align}
Using these notations,
we can write $L_\alpha^n$ as
\begin{align}\label{Lrf}
L_\alpha^n(Q)=p_n(\alpha)\Phi_n(\ana^* Q \ana),\quad Q\in \alm{-1}\otimes{\cal B},\quad n\ge 2.
\end{align}

Next, note that for $R\in \al_{[-n+1,-1]}\otimes\cb,\;n\ge 2$,
an element
\begin{align}\label{br}
\lmk
\tau_{c-}\circ(id_{(-\infty,-2]}\otimes{\cal E})
\rmk^{n-1}
\lmk
e^{\frac\alpha 2 H_\Phi[-n+1,-1]}R
e^{\frac\alpha 2 H_\Phi[-n+1,-1]}
\rmk
\end{align}
belongs to $1_{\alm{-1}}\otimes{\cal B}$,
and (identifying $1_{\alm{-1}}\otimes{\cal B}$
with $\cb$),
\[
\rho\lmk
\lmk
\tau_{c-}\circ(id_{(-\infty,-2]}\otimes{\cal E})
\rmk^{n-1}
\lmk
e^{\alpha  H_\Phi[-n+1,-1]}
\rmk\rmk
=\omega(e^{\alpha  H_\Phi[-n+1,-1]})=p_{n}(\alpha).
\]
Accordingly, 
\[
\varphi_n(R):=
p_{n}(\alpha)^{-1}
\rho\lmk
\lmk
\tau_{c-}\circ(id_{(-\infty,-2]}\otimes{\cal E})
\rmk^{n-1}
\lmk
e^{\frac\alpha 2 H_\Phi[-n+1,-1]}R
e^{\frac\alpha 2 H_\Phi[-n+1,-1]}
\rmk\rmk
\]
defines a state on $\al_{[-n+1,-1]}\otimes\cb$.
We claim
\begin{align}\label{sbd}
s^{-1}\varphi_n(R)
\le
\Phi_n(R)
\le
s\varphi_n(R),\quad\forall R\ge 0,\quad R\in\al_{[-n+1,-1]}\otimes\cb.
\end{align}
To see this, we denote (\ref{br}) by $1_{\alm{-1}}\otimes b_R$.
We have
\begin{align*}
&\Phi_n(R)=
p_n^{-1}(\alpha)
\lmk
\tau_{c-}\circ(id_{(-\infty,-2]}\otimes{\cal E})
\rmk
\lmk
\lmk
\tau_{c-}\circ(id_{(-\infty,-2]}\otimes{\cal E})
\rmk^{n-1}
\lmk
e^{\frac\alpha 2 H_\Phi[-n+1,-1]}R
e^{\frac\alpha 2 H_\Phi[-n+1,-1]}
\rmk
\rmk\\
&=p_n^{-1}(\alpha)\lmk
1_{\alm{-1}}\otimes
\hat{\cal E}_1(b_R)\rmk.
\end{align*}
Therefore, from the bound (\ref{cpb}),
we obtain the claim:
\begin{align}
s^{-1}\varphi_n(R)
=s^{-1}p_n^{-1}(\alpha)\rho(b_R)
\le
\Phi_n(R)=p_n^{-1}(\alpha)
\lmk
1_{\alm{-1}}\otimes
\hat{\cal E}_1(b_R)\rmk
\le
sp_n^{-1}(\alpha)\rho(b_R)=
s\varphi_n(R).
\end{align}
From (\ref{sbd}), we have
$0\le \Phi_n(1)\le s$.
As $\Phi_n$ is completely positive,
we obtain $\lV\Phi_n\rV=\lV\Phi_n(1)\rV\le s$.

We now check the condition (iii).
As in Section \ref{gibbs}, 
there exists a positive constant $C>0$ such that
\begin{align}\label{anbdf}
\sup_{n\in{\mathbb N}}\lV
\ana
\rV,\quad \sup_{n\in{\mathbb N}}\lV
\ana^{-1}
\rV<C.
\end{align}
Furthermore, we have
\begin{align*}
\lim_{n\to\infty}
\lV
[Q, \ana]
\rV=0,\quad \forall Q\in\al_{loc}.
\end{align*}
For a strictly positive element
$Q$
in $\al_{[-n_0,-1]}\otimes\cb$,
we can choose $\varepsilon>0$ and $N(Q)\in{\mathbb N}$ so that
\[
2\varepsilon\lV Q^{\frac 12}\rV C\le
\frac{1}{2C^2} s^{-2}\inf Q,
\]
and
\begin{align*}
n_0+1 \le N(Q),\quad
\lV[ Q^{\frac 12},\ana ]\rV<\varepsilon,\; \forall n\ge N(Q).
\end{align*}
Thus, due to 
the inequality (\ref{sbd}),
for $n\ge N(Q)$,
we have
\begin{align*}
&L_\alpha^n(Q)=
p_n(\alpha)\Phi_n(\ana^* Q\ana)
\le
2C\lV\Phi_n\rV
\lV[Q^{\frac 12},\ana]\rV
\lV Q^{\frac 12}\rV
p_n(\alpha)
+C^2s  p_n(\alpha)\varphi_n(Q)\\
&\le
p_n(\alpha)
\lmk
\frac{1}{2C^2}s^{-1}
+C^2 s
\rmk\varphi_n(Q),\\
&L_\alpha^n(Q)
\ge
-2\lV\Phi_n\rV C\lV[Q^{\frac 12},\ana]\rV
\lV Q^{\frac 12}\rV
p_n(\alpha)
+\frac{1}{C^2}s^{-1}
\varphi_n(Q)p_n(\alpha)
\ge
p_n(\alpha)
\varphi_n(Q)
\frac{1}{2C^2}s^{-1}.
\end{align*}
Hence for $n\ge N(Q)$, we obtain
\begin{align*}
L_\alpha^n(Q)\le
2C^2 s
\lmk
\frac{1}{2C^2}s^{-1}
+C^2 s
\rmk
\inf L_\alpha^n(Q).
\end{align*}
We thus showed (iii).
$\square$\\\\
{\it Proof of Theorem \ref{tfc}}\\
Note that 
the map
\[
{\mathbb R}\ni \alpha \mapsto L_\alpha\in B(F_\theta)
\]
has a $B(F_\theta)$-valued analytic extension to a neighborhood
of $\mathbb R$.
We thus can apply the left-side version of 
Theorem \ref{spec}
to $\{L_\alpha\}$, and obtain
\[
\lim_{n\to\infty}\lV
\lambda(\alpha)^{-n}L_\alpha^n(1)-h(\alpha)
\rV=0,
\]
for some strictly positive element 
$h(\alpha)$ in $\alm{-1}\otimes{\cal B}$ and 
a strictly positive constant $\lambda(\alpha)$.
Furthermore, $\lambda(\alpha)$ is differentiable with respect to
$\alpha$.
By (\ref{Lrf}), (\ref{sbd})and
(\ref{anbdf}),
we have
\begin{align}
\frac{1}{sC^2}p_n(\alpha)\le
C^{-2}p_n(\alpha)\Phi_n(1)
\le
L_\alpha^n(1)=p_n(\alpha)\Phi_n(\ana^*\ana)
\le
C^2 p_n(\alpha)\Phi_n(1)
\le
C^2s p_n(\alpha).
\end{align}
For any state $\nu$ on $\alm{-1}\otimes{\cal B}$,
we obtain
\begin{align}
\lim_{n\to\infty}\frac 1n \log \omega \lmk
e^{\alpha H_{\Phi}[-n,-1]}
\rmk
=\lim_{n\to\infty}\frac 1n \log 
p_n(\alpha)
=\lim_{n\to\infty}\frac 1n \log
\nu \lmk
{L_\alpha}^n(1)\rmk
=\log \lambda(\alpha).
\end{align}
As $\log\lambda(\alpha)$ is differentiable, we have
proved the Theorem.
$\square$
\section{Equivalence of Ensembles}\label{eqe}
An immediate consequence of Theorem \ref{tkms} is 
the equivalence of ensembles considered in \cite{ee}.
Let $\Phi_1,\cdots, \Phi_K$ be translation invariant 
finite range interactions and $X_{1,N},\cdots, X_{K,N}$
corresponding macroscopic observables:
$
X_{k,N}:=\frac{1}{N}H_{\Phi_{k}}[1,N]
$.
Several notions of concentration of macroscopic observables
were introduced in \cite{ee}:\\
A sequence of projections
 $\{P_N\}_{N},\; P_N\in{\mathfrak A}_{[1,N]}$,
is said to be concentrating at 
$x\in{\mathbb R}^K$ whenever
\[
\lim_{N\to\infty} 
\frac{Tr_{[1,N]}\lmk F(X_{k,N}) P_N\rmk}{Tr_{[1,N]}\lmk P_N\rmk}
=F(x_k),
\]
for all $F\in C({\mathbb R})$ and $k=1,\cdots K$,
and written $P_N{\overset{mc}\rightarrow}x$.
In order to define concentration of states,
we need a set $\cal F$
of maps $G$ from a set of all finite sequence of
$\{1,\cdots, K\}$, $I$, to $\mathbb C$,
such that
\[
\sum_{m\ge 0}\sum_{(k_1,\cdots,k_m)\in I}
\lv
G(k_1,\cdots,k_m)
\rv
\prod_{i=1}^{m}\lV\Phi_{k_i}\rV<\infty.
\] 
We define $G(X^N)$ by
\[
G(X^N):=\sum_{m\ge 0}\sum_{(k_1,\cdots,k_m)\in I}
G(k_1,\cdots,k_m)X_{k_1,N}\cdots X_{k_m,N}.
\]
A sequence of states $\omega_N$ on ${\mathfrak A}_{[1,N]}$,
is concentrating at $x\in{\mathbb R}^K$ if
\[
\lim_{N\to\infty}\omega^N(G(X^N))=G(x),
\] 
for all $G\in{\cal F}$, and written
$\omega^N\rightarrow x$.
It was shown in \cite{ee} that if $P_N\overset{mc}\rightarrow x$,
then the states $\frac{Tr_{[1,N]}\lmk \cdot  P_N\rmk}{Tr_{[1,N]}\lmk P_N\rmk}
\rightarrow x$.
Furthermore, we write
$\omega^N\overset{1}\rightarrow x$ whenever
$\lim_{N\to\infty} \omega^N(X_{k,N})=x_k$.
Three H-functions $H^{mc}$, $H^{can}$, $H^{can}_1$
were introduced in \cite{ee}:
\begin{align*}
H^{mc}(x):=\sup_{P^N\overset{mc}\rightarrow x}\limsup_{N\to+\infty}\frac{1}{N}
\log Tr_{[1,N]}(P^N),\\
H^{can}(x):=\sup_{\omega^N\to x}\limsup_{N\to+\infty}
\frac{1}{N}{\cal H}(\omega^N),\\
H^{can}_1(x):=\sup_{\omega^N\overset{1}\rightarrow x}\limsup_{N\to+\infty}
\frac{1}{N}{\cal H}(\omega^N),
\end{align*}
where ${\cal H}(\omega^N)$ is the von Neumann entropy of $\omega^N$.
By definition, we have $H^{mc}(x)\le H^{can}(x)\le H^{can}_1(x)$. 
The following theorem was proven in \cite{ee}.
\begin{thm}\label{dmn}
Assume that there exists a sequence of states $\omega^N$ on 
${\mathfrak A}_{[1,N]}$ with density matrices $\sigma^N$,
satisfying the following conditions:
\begin{description}
\item[(i)]
For all $\delta>0$ and $k$, there exists $C_k(\delta)>0$ and 
$N_k(\delta)\in{\mathbb N}$ suth that
\begin{align*}
\int_{x_k-\delta}^{x_k+\delta}\omega^N(Q_N^k(d\lambda))
\ge 1-e^{-C_k(\delta)N},\quad \forall N\ge N_k(\delta),
\end{align*}
where $Q_N^k$ is the spectral projection of $X_{k,N}$.
\item[(ii)]
For all $\delta>0$,
\begin{align*}
\lim_{N\to\infty}\frac{1}{N}\log\int_{-\delta}^\delta
\omega^{N}(\tilde Q_{N}(d\lambda))=0,
\end{align*}
where $\tilde Q_N$ is the spectral projection of 
$\frac{1}{N}(\log\sigma_N-Tr_{[1,N]}\sigma_N\log \sigma_N)$.
\item[(iii)]
$H_1^{can}(x)=\lim_{N\to\infty}\frac{1}{N}{\cal H}(\omega^N)$.
\end{description}
Then we have
\begin{align*}
H^{mc}(x)=H^{can}(x)=H^{can}_1(x).
\end{align*}
\end{thm}
\quad \\

This means the equivalence of microcanonical
ensemble and canonical ensemble.
Let us consider 
a sequence of states of the form 
\begin{align}\label{form}
\omega^N(A)=
\frac{Tr_{[1,N]} e^{\sum_k\lambda_kH_{\Phi_k}[1,N]} A}{Tr_{[1,N]}e^{\sum_k\lambda_kH_{\Phi_k}[1,N]}},\quad \lambda_k\in{\mathbb R}.
\end{align}
Theorem \ref{tkms} and a bound similar to (\ref{st}) guarantee that
$\omega^N$ concentrates at $x$ for some $x\in{\mathbb R}^K$ and
satisfies conditions (i),(ii) of Theorem \ref{dmn}.
Furthermore, 
it can be shown that a state of this form satisfies (iii)
\cite{ee}.
Therefore, applying Theorem \ref{dmn},
we obtain the equivalence of ensembles
in one dimensional quantum spin system:
\begin{cor}
If there exists a sequence of states of $\omega^N$ of the form
(\ref{form}) such that $\omega^N\overset{1}\rightarrow x\in{\mathbb R}^K$, then
\[
H^{mc}(x)=H^{can}(x)=H^{can}_1(x).
\]
\end{cor}

\appendix
\quad\\
\noindent
{\bf Acknowledgement.}\\
{The author thanks for Professor L.Rey-Bellet
and Dr. W. De Roeck for interesting discussions.
}
\noindent

\section{Analyticity of local elements}\label{anal}
Let $I$ be any subset of $\mathbb Z$ and $\Phi$
a finite range interaction.
We define a new interaction $\Phi_I$ by
\begin{align*}
\Phi_I(X):=\left\{
\begin{gathered}
\Phi(X),\quad if\; X\subset I\\
0,\quad otherwise
\end{gathered}
\right. .
\end{align*}
This new interaction gives a time evolution $\tau_{\Phi_I}$.
We define $\al_1$ by
\begin{align}\label{a1}
\al_1:=\left\{
Q\in F_\theta\cap\ali{1}:0<\forall \theta<1
\right\}.
\end{align}
In \cite{AraGib}, H.Araki showed that
 $\al_1$ is a $*-$ algebra and
that any element in $\al_1$ 
is entire analytic for $\tau_{\Phi_I}$.
For a local element $Q$, we define $E_r(Q; H_\Phi(I))$ by
\[
E_r(Q; H_\Phi(I))\equiv \sum_{n=0}^\infty
\int_0^1d\beta_1\int_0^{\beta_1}d\beta_2\cdots
\int_0^{\beta_{n-1}}d\beta_n
\tau_{\Phi_I}^{-i\beta_n}(Q)\cdots
\tau_{\Phi_I}^{-i\beta_1}(Q).
\]
It was shown in \cite {AraGib} that $E_r(Q; H_\Phi(I))$
is an element in $\al_1$.
Furthermore,
following relations hold:
\begin{align}\label{com}
E_r(Q_1+Q_2; H_\Phi(I))=E_r(Q_1;Q_2+H_\Phi(I))E_r(Q_2;H_\Phi(I)),
\nonumber\\
E_r(Q;H_\Phi(I))\tau_{\Phi_I}^{-i}(Q')
=\tau_{\Phi_I+Q}^{-i}(Q')E_r(Q;H_\Phi(I)),
\end{align}
for all $Q_1,Q_2,Q\in\al_{loc}$ and $Q'\in \al_1$.
Here, $\tau_{\Phi_I+Q}$ is a perturbed
dynamics of $\tau_{\Phi_I}$
by a bounded perturbation $Q$.
If $Q\in \al_{loc}$,
then for any $x>1$, there exists a constant $C_x$ 
such that 
\begin{align*}
\sup_{N\in{\mathbb N}}
x^N\cdot \lV E_r(Q;H_\Phi(I))-E_r(Q;H_\Phi(I\cap [-N,+N]))\rV
\le C_x.
\end{align*}
We use the following notations:
\begin{align*}
&\hat H^r_\Phi(n):=
\sum_{I\subset[1,\infty),I\cap[1,n]\neq\phi}\Phi(I)
\quad\in \ali{1}\cap \al_{loc},\\
&\hat H^l_\Phi(n):=
\sum_{I\subset(-\infty,-1],I\cap[-n,-1]\neq\phi}\Phi(I)
\quad\in \alm{-1}\cap \al_{loc},\\
&W^r_\Phi(n):=\sum_{
I\subset [1,\infty),I\not\subset [1,n-1]
,I\not\subset [n+1,\infty)}\Phi(I)
\quad \in \ali{1}\cap \al_{loc},
\\
&W^l_\Phi(n):=\sum_{I\subset (-\infty,-1],I\not\subset
[-n+1,-1],
I\not\subset(-\infty,-n-1]
}\Phi(I)\quad\in\alm{-1}\cap \al_{loc}.
\end{align*}
We may apply the same argument as \cite{AraGib}
to show the
following facts:
\begin{lem}\label{ks}
Let $\Phi$ and $\Psi$ be finite range interactions
with range less than $r>0$.
Then operators
\begin{align*}
\ana:=&
\tau_{\Phi_{[1,\infty)}}^{-i\frac{\alpha}{2}}\lmk
E_r\lmk -\frac{\beta}{2} W^r_\Psi(n); -\frac{\beta}{2} 
(H_\Psi[1,n]+H_\Psi[n+1,\infty))\rmk\rmk\\
&\cdot E_r\lmk
\frac{\alpha}{2} W_\Phi^r(n); \frac{\alpha}{2} (H_\Phi[1,n]
+H_\Phi[n+1,\infty))\rmk,\nonumber\\
a_n^{N}(\alpha):=&
\tau_{\Phi_{[n-N,n+N]\cap{[1,\infty)}}}^{-i\frac{\alpha}{2}}\lmk
E_r\lmk -\frac{\beta}{2} W_\Psi^r(n); -\frac{\beta}{2} 
(H_\Psi([n-N,n]\cap[1,\infty))+H_\Psi([n+1,n+N])\rmk\rmk\\
&\cdot E_r\lmk
\frac{\alpha}{2} W_\Phi^r(n);
 \frac{\alpha}{2} (H_\Phi([n-N,n]\cap[1,\infty))+
H_\Phi[n+1,n+N)\rmk,\\
\alpha,\beta\in {\mathbb C},&\quad n\in{\mathbb N}
\end{align*}
are well-defined invertible
 elements in $\al_1$
and 
$\al_{[n-N-r,n+N+r]\cap[1,\infty)}$,
respectively.
For any compact set $S$ in $\mathbb C$,
there exists a positive constant $C_S$ such that
\begin{align*}
&\sup_{\alpha\in S}
\sup_{n\in {\mathbb N}}
\lV \ana\rV,\;
\sup_{\alpha\in S}
\sup_{n\in {\mathbb N}}
\lV \lmk \ana\rmk^{-1}\rV<C_S,\\
&\sup_{N\in{\mathbb N}}\sup_{\alpha\in S}
\sup_{n\in {\mathbb N}}
\lV a_n^{N}(\alpha)\rV,\;
\sup_{N\in {\mathbb N}}\sup_{\alpha\in S}
\sup_{n\in {\mathbb N}}
\lV \lmk a_n^{N}(\alpha)\rmk^{-1}\rV<C_S.
\end{align*}
Furthermore, 
for any $x>1$, there exists a positive constant $C_x$ 
such that 
\begin{align*}
\sup_{N\in{\mathbb N}}\sup_{\alpha\in S}
\sup_{n\in{\mathbb N}}
x^N\cdot 
\lV 
\ana-a_n^{N}(\alpha)
\rV
\le C_x.
\end{align*}
\end{lem}


\end{document}